\def\simless{\mathbin{\lower 3pt\hbox
             {$\rlap{\raise 5pt\hbox{$\char'074$}}\mathchar"7218$}}}    

\def\simmore{\mathbin{\lower 3pt\hbox
             {$\rlap{\raise 5pt\hbox{$\char'076$}}\mathchar"7218$}}}    

\documentclass[structabstract]{aa}

\usepackage{graphicx}
\begin{document}
\title{
Spectral formation in a radiative shock:  
application to anomalous X-ray pulsars and soft gamma-ray repeaters
}

\subtitle{}

\author{N. D. Kylafis\inst{1,2}, J. E. Tr\"{u}mper\inst{3}, \and \"U. Ertan\inst{4}}

\institute{
University of Crete, Physics Department \& Institute of
Theoretical \& Computational Physics, 71003 Heraklion, Crete, Greece\\
\and
Foundation for Research and Technology-Hellas, 71110 Heraklion, Crete, Greece\\
\and
Max-Planck-Institut f\"{u}r extraterrestrische Physik, 
Postfach 1312, 85741 Garching, Germany\\
\and
Faculty of Engineering and Natural Sciences, Sabanc\i\ University,
34956, Orhanl\i, Tuzla, \.Istanbul, Turkey
}

\date {Received ; accepted }


\abstract 
{
In the fallback disk model for the persistent emission of Anomalous
X-ray pulsars (AXPs) and soft gamma-ray repeaters (SGRs), the hard X-ray
emission arises from bulk- and thermal Comptonization of bremsstrahlung
photons, which are generated in the accretion column.  The relatively
low X-ray luminosity of these sources implies a moderate transverse
optical depth to electron scattering, with photons executing a small
number of shock crossings before escaping sideways.
}
{
We explore the range of spectral shapes that can be obtained with this
model and characterize the most important parameter dependencies.
}
{We use a Monte Carlo code to study the crisscrossing of photons in a
radiative shock in an accretion column
and compute the resulting spectrum.}
{As expected, high-energy power-law X-ray spectra are produced in
radiative shocks with photon-number spectral index $\Gamma \simmore 0.5$.
We find that the required transverse optical depth is 
$1 \simless \tau_{\perp} \simless 7$.
Such spectra are observed in low-luminosity
X-ray pulsars.}
{We demonstrate here with a simple model
that Compton upscattering in 
the radiative shock in the accretion column 
can produce hard X-ray spectra similar to those seen
in the persistent and transient emission of AXPs and SGRs.
In particular, one can obtain a high-energy power-law spectrum, with
photon-number spectral-index $\Gamma \sim 1$ 
and a cutoff at $100 - 200$ keV, 
with a transverse Thomson optical depth of $\sim 5$,
which is shown to be typical in AXPs/SGRs.}


\authorrunning{Kylafis \& Tr\"{u}mper 2013}
\titlerunning{Spectral formation in radiative shocks} 

\maketitle


\section{Introduction}

Anomalous X-ray Pulsars (AXPs) and Soft Gamma-ray Repeaters (SGRs) are 
young neutron stars that have been observed to emit X-rays both 
quiescently and in the form of bursts.  They are commonly called 
{\it magnetars}, because they are thought to have super-strong magnetic
fields ($10^{14} - 10^{15}$ G).  
For recent reviews see Woods \& Thompson (2006) and Mereghetti (2008, 2013).
Two main models have been proposed to explain the observational data;
the classical magnetar model (e.g., Thompson \& Duncan 1993, 1996) 
and the fallback disk one (Chatterjee et al. 2000; Alpar 2001).
Both models have
in common that the giant and the highly super-Eddington bursts are produced
by magnetic field decay.  The fallback disk model puts the
super-strong magnetic field in a multipole component, while 
the classical magnetar model puts it in the dipole component.  Yet, a
recent interpretation (Tiengo et al. 2013) of a variable absorption
feature in the spectrum of the magnetar SGR 0418+5729 as a proton 
cyclotron feature invokes a multipole magnetic field.  Thus, 
multipole components could in principle operate in both models.

For the persistent emission,
the explanation is different in the two models.  The classical
magnetar model assumes that the persistent emission is  
powered by magnetic field decay and that
the observed period derivative $\dot P$ is due to magnetic braking of
an isolated neutron star with a super-strong dipole magnetic field.
On the other hand, in the fallback disk model 
the persistent X-ray emission is powered by accretion onto the neutron star,
the optical and the infrared radiation is produced by the disk, 
and the observed rotational properties are explained by the long-term
disk - magnetosphere interaction.  The evolution of AXPs and SGRs
with fallback disks has been 
studied in a number of papers (e.g., Ertan et al. 2007, 2009) and 
a fallback disk has been observed in two magnetars: AXP 4U 0142+61
(Wang et al. 2006) and AXP 1E 2259+586 (Kaplan et al. 2009). 

A detailed model for the formation of the quiescent spectra of AXPs/SGRs 
has been proposed by Tr\"umper et al. (2010, 2013)
using the fallback disk model.  These spectra consist 
of two components (e.g., den Hartog et al. 2008a,b; Enoto et al. 2010); 
a quasi-thermal component at low energies, which is generally attributed 
to the photospheric emission from the neutron star polar cap, and a 
hard power-law component. In the model of Tr\"umper et al. (2010, 2013), 
this power-law component is attributed to the emission from the radiative 
shock, located near the bottom of the accretion column. 
This component is emitted as a fan beam, which partly hits the 
surrounding polar cap, where the radiation is absorbed or scattered, 
leading to the quasi-thermal component. The geometry  of the model
ensures that both soft and hard components have approximately the same 
luminosity (as observed), while the observed luminosity ratio depends 
on the height of the radiative shock above the neutron-star surface 
and the angles between the line of sight on one hand
and the magnetic-dipole and the spin axis on the other. 

In this paper we concentrate specifically on the formation of the fan 
beam in the radiative shock and explore the conditions required to 
produce a hard power-law spectrum.  Our picture is similar to that
of X-ray pulsars.

It is well established that in X-ray pulsars matter from a companion star
falls onto the magnetic poles of a neutron star.  The accreting matter 
follows dipole 
magnetic-field lines and falls supersonically.  Eventually, it has to
stop on the neutron-star surface.  The supersonic to subsonic transition
of the accretion flow 
occurs in a radiative shock above the neutron-star surface (Basko \& Sunyaev
1976).  
In such a shock, which for simplicity we consider here as a
mathematical discontinuity,
soft X-ray photons coming mainly from below the shock
suffer a nearly head-on collision with the infalling electrons.  
Because of the large momentum of the infalling eletrons with respect 
to that of the soft photons, the photons are scattered backward,
gain energy, and return to the dense, hot, subsonic flow.  
We will refer to this as bulk-motion Comptonization (BMC).
In the subsonic flow, the photons are scattered again, more or less
isotropically.  If their energy is significantly less
than the thermal energy of the electrons, the
photons gain energy on average (thermal Comptonization, hereafter TC).
After the scattering, some photons find themselves again in the supersonic
flow and the above processes are repeated, albeit with progressively
decreasing probability, because the photons can escape sideways, i.e.,
perpendicular to the accretion column.  We want to remark here that
in a radiative shock, like the one we envision above, it is impossible
to separate BMC from TC, contrary to what happens in a spherical flow,
where one can have pure BMC (Blandford \& Payne 1981a,b; Payne \&
Blandford 1981; Mastichiadis \& Kylafis 1992; Titarchuk et al. 1996, 1997).
  
It is well known that this first-order
Fermi energization of the photons results in a power-law,
high-energy spectrum.
Lyubarski \& Sunyaev (1982, hereafter LS82) were the first to compute
analytically the emergent spectrum from such a shock.  They studied
spectral formation due to BMC and TC in a radiative shock that is
of the Rankine-Hugoniot form.  This is a reasonable approximation, 
given the fact that, due to the large optical depth along 
the accretion column, the upscattered photons escape sideways. 

Accretion onto magnetic neutron stars has been studied extensively by
Becker \& Wolff.  In a series of papers (Becker \& Wolff 2005a; 2005b,
2007; Becker et al. 2012), 
they have studied spectral formation in X-ray pulsars and have
highlighted the role of BMC and TC for spectral formation 
in accretion columns.  In their classic paper (Becker \& Wolff 2007,
hereafter BW07), they consider a smooth flow in a cylindrical
column, with the flow being slowed down by radiation.  The
shock is treated properly, not as a mathematical discontinuity,
and it is the place where BMC and TC operate with seed soft photons
produced by bremsstrahlung, cyclotron, and blackbody processes inside 
the column.  As most of the photons escape sideways, the model
predicts a fan beam.

In X-ray pulsars, the accretion column has a large optical depth in all
directions.  This means that the radiation in the column thermalizes
to a significant degree 
and any high-energy photons present in the column
are downscattered.  Thus, the emergent 
spectrum has a cutoff at $\sim 20$ keV (BW07).  In low-luminosity
pulsars, on the other hand, the sideways optical depth is expected
to be of order a few.  Thus, a high-energy power-law spectrum with
a much larger cutoff energy can in principle be produced.  This is what we 
investigate here.

In \S\ 2 we describe the model that we have used, giving a qualitative
description of the processes that occur in the radiative shock,
in \S\ 3 we present our results, in \S\ 4 we discuss them,
and in \S\ 5 we draw our conclusions and state the predictions of our work.

\section{The model}

In order to demonstrate our main ideas, we use a relatively simple model.
We consider a cylindrical accretion column with two regions:  the pre-shock
region, where matter falls  
with constant velocity, and the post-shock one, which has 
density seven times the pre-shock one, electron temperature $T_e$, and
negligible flow velocity.  To fit real data, significantly more
elaborate models are needed.

Mainly bremsstrahlung photons from below the shock,
but also soft X-ray photons from 
the neutron-star photoshpere surrounding the accretion
column, enter the shock
and are followed by a Monte Carlo code until they escape.  Since the
Thomson optical depth along the accretion column in the pre-shock 
region is infinite by construction in our model, the crucial 
parameter in our calculation is the Thomson optical depth in the 
transverse direction in this region. 

\subsection{Qualitative presentation}

\subsubsection{The radiative shock}

If for simplicity 
we consider that the accreting plasma consists only of protons 
and electrons, the accretional energy resides in the protons.  Charge 
neutrality guarantees that the electrons are tied to the protons.  Thus,
the electrons have the same free-fall velocity as the protons, but
insignificant accretional energy.  The accretional energy of the infalling
plasma per proton-electron pair is
$$
E_p = { {GMm_p} \over R },
\eqno(1)
$$
where $G$ is the gravitational constant, $M$ is the mass of the neutron star,
$R$ is its radius, and $m_p$ is the proton mass.
For a neutron star of mass $M=1.4 M_{\odot}$ and
radius $R=12.5$ km, $E_p \approx 150$ MeV. 

If a significant amount of this energy is to be transfered to the photons via
BMC, it looks, at first sight, impossible to happen in the radiative
shock. This is because the energy $E_{\gamma}$ 
of a soft photon becomes after a head-on 
collision with a relativistic electron 
$$
E_{\gamma}^{\prime} \approx { {1+v/c} \over {1-v/c} } E_{\gamma} 
\approx 4 E_{\gamma},
\eqno(2)
$$
for an electron speed of, say, $v = 0.6 c$.  This energy gain by the photon
is orders of magnitude smaller than $E_p$.
As we demonstrate below, though, it is possible for the photons to
``steal'' a significant fraction of the accretional energy,
because many photons scatter off the same electron.

For simplicity, 
we think of the shock as a mathematical discontinuity, but in reality it has
a finite thickness with characteristic scale equal to the photon
mean free path for electron scattering
$$
\bar l_e ={1 \over {n_e \sigma_T} },
\eqno(3)
$$
where $n_e$ is the electron number density in the shock and 
$\sigma_T$ is the Thomson cross section.  For simplicity, in this
discussion, we use the Thomson cross section, but in our Monte
Carlo calculation we use the angle- and energy-dependent one,
due to the magnetic field (see \S\ 2.2).

In this characteristic thickness $\bar l_e$, the ``optical depth'' that
an infalling electron sees in the ``bath'' of outgoing photons is
$$
\tau_{\gamma}=n_{\gamma} \sigma_T \bar l_e= 
{{\bar l_e} \over {\bar l_{\gamma}} } =
{n_{\gamma} \over n_e},
\eqno(4)
$$
where $n_{\gamma}$ is the number density of photons that mediate the
shock and $\bar l_{\gamma}$ is the electron mean free path in the
``bath'' of photons.
For the electron number density we write
$$
n_e= {{\dot M} \over {m_p \pi a_0^2 v_{ff}} },
\eqno(5)
$$
where $\dot M$ is the mass accretion rate,
$a_0$ is the radius of the accretion column at the shock, and
$v_{ff}=\sqrt{2GM/R} = 0.58 c$ is the free-fall velocity of the plasma
near the neutron star surface, where the shock is formed.  Here we have 
taken $M= 1.4 M_{\odot}$ and $R=12.5$ km.

Let $u_{\gamma}$ be the energy density of the photons in the shock and 
$E_{\gamma}$ their characteristic energy.  Then, the accretional
luminosity can be written as
$$
L=u_{\gamma} c \pi a_0^2 = n_{\gamma} E_{\gamma} c \pi a_0^2 
= { {GM \dot M} \over R },
\eqno(6)
$$
from which we get
$$
n_{\gamma}= {1 \over {E_{\gamma} c \pi a_0^2} } { {GM \dot M} \over R}.
\eqno(7)
$$
Thus, the ``optical depth'' $\tau_{\gamma}$ becomes
$$
\tau_{\gamma} = { {v_{ff}} \over c}
{ {GMm_p/R} \over {E_{\gamma}}} ={ {v_{ff}} \over c} {E_p \over E_{\gamma} }
\sim {E_p \over E_{\gamma}},
\eqno(8)
$$
which implies that the ``optical depth'' that an infalling electron sees
in the ``bath'' of outgoing photons in the shock is of order the ratio of the
accretional energy of a proton-electron pair to the typical energy
of the outgoing photons.  This means that, in the thickness of the shock,
many photons scatter off the same electron and this is how the kinetic
energy of the accreting plasma is reduced in the shock.
As an analogy, one could think of the stopping of a diver by the water 
molecules in a pool.
Thus, we envision the transfer of energy from the protons to the electrons 
and from them to the photons as follows:

At the ``top'' of the shock, consider a head-on collision between an 
outgoing photon and an infalling electron.  The energy gain of the photon
is described by eq. (2) and this energy is taken from the electron.  Thus,
the electron would slow down, and charge separation would occur,
if it were not for the protons, which
transfer a tiny fraction $\sim E_{\gamma} / E_p$ of their energy
to replenish the lost energy of the electron. This process is repeated
many times as the proton-electron pairs traverse the shock thickness.
Thus, at the ``bottom'' of the shock, the protons have given up 
a significant fraction of their accretional energy to the photons
that mediated the shock, with the electrons acting as intermediaries.

\subsubsection{The postshock temperature}

According to eq. (2), the energy of a soft photon quadruples in a head-on
collision with an infalling electron.  Therefore, after several
crisscrosses of the shock, the energy of a photon increases significantly,
but, at the same time, the photon looses some of its energy due to 
electron recoil in the postshock region.  This energy is deposited in
the postshock region and heats the electrons (LS82).

An estimate of the electron (and proton) temperature in the postshock
region can be obtained as follows:

The energy lost to the photons via BMC, is replenished by the accretional 
energy of the protons.  This channeling of energy stops when the kinetic
energy of the protons becomes comparable to the accretional energy of
the electrons $E_e=GMm_e/R \approx 75$ keV, 
where $m_e$ is the electron mass.  Thus, a
rough estimate of the electron temperature $T_e$ in the postshock region is
given by
$$
{ {GMm_e} \over R} \sim {1 \over 2} kT_e,
\eqno(9)
$$
where, due to the strong magnetic field, we have assumed a one-dimensional
Maxwellian distribution.  This gives $kT_e \sim 150$ keV.
In our calculations, and for completeness, we have considered a range of
temperatures.

\subsubsection{The size of the accretion column}

In what follows, we estimate the radius of the accretion column at the
radiative shock, close to the neutron-star surface.

For a neutron star with surface magnetic-dipole field $B$ at the equator  
and magnetic moment $\mu=BR^3$, the Alfven radius, to within a factor
of order unity, is (Lamb et al. 1973; Frank et al. 2002)
$$
r_A = (GM)^{-1/7} \mu^{4/7} {\dot M}^{-2/7}.
\eqno(10)
$$
The fallback disk,
which is assumed to be of the Shakura \& Sunyaev (1973) type, 
has a pressure scale height (Frank et al. 2002)
$$
h=1.2 \times 10^8 \alpha^{-1/10} {\dot M}_{15}^{3/20} 
\left( {M \over {M_{\odot}}} \right)^{-3/8} r_{10}^{9/8} f^{3/5}~{\rm cm},
\eqno(11)
$$
where $\alpha = 0.03$ is the viscosity parameter,
${\dot M}_{15}$ is the accretion rate in units of $10^{15}$ g s$^{-1}$, 
$r_{10}$ is the radial distance in units of $10^{10}$ cm
and
$$
f=[1-(R/r)^{1/2}]^{1/4} \approx 1.
$$
Assuming that the inner disk couples to the magnetic-dipole field lines
in an annular region of radius $r_A$ and radial thickness $\Delta r$, the
cross-sectional area ($\pi a_0^2$) of the accretion column
near the surface of the neutron star
can be estimated from magnetic-flux
conservation, namely
$$
B(r_A) 2 \pi r_A \Delta r = \pi a_0^2 B_p ,
\eqno(12)
$$
where $B(r_A)=(1/2) B_p (R/r_A)^3$ and $B_p = 2 B$ is the strength
of the magnetic-dipole field at the pole of the star.

Assuming $\Delta r \approx \eta h$, where $\eta$ is a scaling factor, 
we find
$$
a_0 \approx 50 \left( {\eta \over 3} \right)^{1/2} 
{\dot M}_{15}^{1/5} B_{12}^{-1/4}~{\rm m}.
\eqno(13)
$$
Alternatively,
if the inner-disk matter couples to the magnetic-dipole field lines 
between $r_A$ and the co-rotation radius $r_{\rm co}=(GM/\Omega^2)^{1/3}$,
where $\Omega$ is the angular frequency of the neutron star, then for
$\Delta r \approx r_A - r_{\rm co} \approx 5 \times 10^8$ cm
we find
$$
a_0 \approx 170~{\dot M}_{15}^{1/5} B_{12}^{-1/4}~{\rm m},
\eqno(14)
$$
which is not significantly larger than what we have found in eq. (13).

The Thomson optical depth $\tau_{\perp}$ in the
direction perpendicular to the magnetic field, 
just above the radiative shock,
is
$$ 
\tau_{\perp}= n_e \sigma_T a_0 = 
{{\dot M} \over {m_p v_{ff} \pi a_0^2}} \sigma_T a_0 
= 7.3 \times 10^{-12} {{\dot M} \over {a_0}}.
\eqno(15)
$$
For $\dot M = 2 \times 10^{15}$ g s$^{-1}$, which corresponds to an
X-ray luminosity of $3.7 \times 10^{35}$ erg s$^{-1}$, 
and $a_0 \approx 50 $ m, we find $\tau_{\perp} \approx 5.8$.

\subsection{The Monte Carlo code}

Our Monte Carlo code is similar to previously used codes in our work 
and it is based on Cashwell \& Everett (1959) and Pozdnyakov et al. (1983).
For simplicity, we assume cylindrical symmetry and a two-zone model: the
pre-shock zone ($z>0$) and the post-shock one ($z<0$).  The shock is 
of the Rakine-Hugoniot form at $z=0$.  The cross section of the cylinder
has radius $a_0$, which corresponds to a transverse Thomson optical depth 
$\tau_{\perp}$ in the pre-shock zone.  
In the pre-shock zone, matter falls freely with constant velocity
$v_{ff}=0.58 c$, appropriate for a 1.4$M_{\odot}$ neutron star with a
radius $R=12.5$ km.

In the post-shock zone, the matter is compressed 
by a factor of seven (appropriate for an adiabatic hydrodynamic shock
with $\gamma = 4/3$), hot, and
with negligible flow velocity.
In \S\ 2.1.2 we estimated the electron temperature behind the shock to be
$kT_e \sim 150$ keV.  As this value is probably an overestimate, 
because we used a one-dimensional Maxwellian distribution, we use 
three values of $kT_e$:  30, 50, and 100 keV.

Photons,  mainly bremsstrahlung from below the shock,
and to a lesser extend photospheric photons from the polar cap,
enter the shock.
Some of them escape unscattered, while the rest are scattered before
escape.  Of the last ones, some are scattered once, others a few times,
and a relatively small number are scattered many times.  It is this last
category of photons, that have crisscrossed the shock
many times, have increased their energy significantly, and have produced the
high-energy power-law spectrum, as it is expected from such a first-order
Fermi energization.

The input bremsstrahlung photons to the radiative shock 
have temperatute $T_e$ and are emitted isotropically upwards.
In other words, we do not perform radiative transfer on the 
isotropically emitted bremsstrahlung photons in the post-shock region, 
but instead assume an upward flux of such photons.

The input quasi-thermal photospheric photons to the radiative
shock are approximated as blackbody ones with temperature of $kT_{BB}=0.5$ keV. 
The polar cap that emits quasi-thermal photons has a radius $r_0
\sim 8$ km in 4U 0142+61 
(White et al. 1996; Israel et al. 1999; Juett et al. 2002).  Therefore, only
a relatively small fraction of the photospheric photons enter the shock
(see \S\ 3.2 below).

Following BW07, we take the approximation of Arons et al. (1987) for 
the magnetic cross section for continuum scattering by electrons,
namely
$$
\sigma(E, \theta)= \sigma(E) \left[ sin^2\theta+k(E)cos^2\theta \right],
\eqno(16)
$$
where $E$ is the photon energy, 
$\sigma(E)$ is the Klein-Nishina cross section,
$\theta$ is the angle between the 
directions of the photon and the magnetic field, and
$$
k(E) \equiv \cases{1,&$E \ge E_c$,\cr
(E/E_c)^2,&$E<E_c$,\cr}
\eqno(17)
$$
where 
$$
E_c={ {heB} \over {2\pi m_e c} }= 11.6 
\left( {B \over {10^{12} {\rm Gauss}}} \right) {\rm keV}
\eqno(18)
$$
is the cyclotron energy.  In our calculations, we consider two indicative
values of the magnetic-field strength, $10^{12}$ G and $10^{13}$ G.

The mean free path to electron scattering 
from every point in the accretion column and in any direction, as well as
the energy change and the new direction of a photon after scattering, 
are computed using the corresponding relativistic expressions
(Pozdnyakov et al. 1983). 
For the thermal distribution of electron velocities we use a 
one-dimensional Maxwellian.
When a photon escapes, both its energy and direction are recorded.  Thus,
we compute direction-dependent spectra as well as angle-averaged spectra.

\section{Results}

In what follows, we present the results of our Monte Carlo calculations 
in the form of $E^2 dN/dE$ in arbitrary units, as a function of $E$ in keV.
The main parameter in our calculations is 
the transverse Thomson optical depth 
$\tau_{\perp}$ (see eq. 15),
which is almost directly proportional to $\dot M$, since $a_0$ is of 
order $100$ m (see eqs. 13 and 14).
For completeness we explore also three values of the postshock
temperature $kT_e= 30,\ 50$ and $100$ keV, 
and two values of the magnetic field strength $B=10^{12}$ and $10^{13}$ G.
For the bremsstrahlung photons the temperature is the assumed
post-shock value of $T_e$, while
for the blackbody, photospheric temperature we take $kT_{BB}=0.5$ keV.

For clarity of presentation, we treat the bremsstrahlung and the blackbody
input photons separately, while in reality their upscattering in the
shock occurs simultaneously (see, however, \S\ 3.3).  
We start with bremsstrahlung input photons.

\subsection{Bremsstrahlung input}

Below the shock, the main emission mechanism is bremsstrahlung (BW07).  Thus,
the spectrum of the emitted photons is given in the Born approximation 
(Greene 1959) by
$$
{ {dN} \over {dE} } \propto G(E) e^{-E/kT_e}
\eqno(19)
$$
where
$$
G(E)=e^{E/2kT_e}K_0(E/2kT_e),
\eqno(20)
$$
is the Gaunt factor and
$K_0$ is the modified Bessel function of the second kind, zeroth order.

For our first calculation we use the following values of the parameters:
$\tau_{\perp}= 5$, $kT_e= 50$
keV, and $B=10^{12}$ G, and refer to them
as the {\it reference values}.

In Fig. 1, we show the input bremsstrahlung spectrum (crosses), 
with the integral of $dN/dE$ normalized to unity,
and the resulting angle-averaged output spectrum
of the Monte Carlo (stars) for the reference values of the parameters.

\begin{figure}
\centering
\includegraphics[angle=-90,width=10cm]{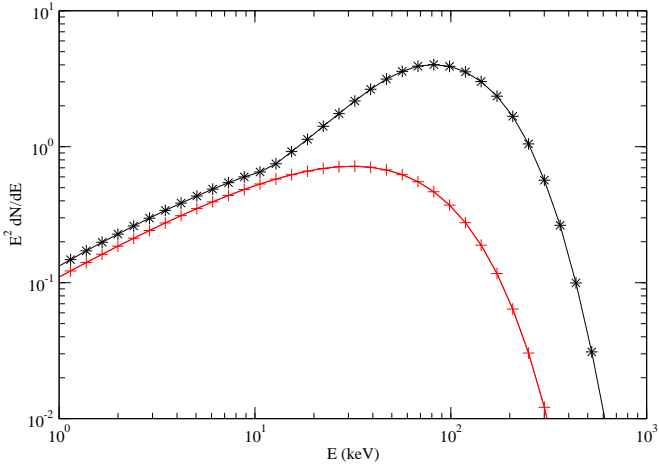}
\caption{
Angle-averaged emergent spectrum (stars) in $E^2 dN/dE$ form, with 
arbitrary units, as a function of energy $E$ in keV.  The spectrum 
of the input photons (normalized to unity) is 
due to bremsstrahlung emission (crosses).  The parameters are at their
reference values.
}
\label{Fig1}
\end{figure}

Above about 10 keV, the spectrum is a power law, with photon-number
spectral index $\Gamma \approx 1$ and a peak in the $E^2 dN/dE$
distribution at $E_p \approx 80$ keV.
This is a natural consequence of the first order Fermi energization.
As it is evident from Fig. 1, most of the energy in the emergent spectrum
comes out at $E_p$.  

Below about 10 keV, the output photon-number spectrum $dN/dE$ is essentially 
the input bremsstrahlung spectrum ``shifted" to higher energy by a 
small amount.  Its
qualitative difference from the spectrum above 10 keV is due to the
change in the cross section at the cyclotron energy $E_c = 11.6$ keV 
(see eq. 18).  Below the cyclotron energy, the cross section is
angle dependent and smaller than the Thomson value (see eqs. 16 and 17),
while above the cyclotron energy it is isotropic and equal to the
Thomson value for low-energy photons.

\subsubsection{Effects of transverse optical depth}

As discussed above, the larger the transverse Thomson optical depth 
$\tau_{\perp}$, the larger the number that a typical photon 
criscrosses the shock.  This has a significant effect on
the emergent spectrum.

In Fig. 2 we show the input bremsstrahlung spectrum (crosses),
normalized to unity, and the 
emergent spectra for four values of $\tau_{\perp}$:  1 (triangles), 
3 (squares), 5 (stars), and 7 (circles). 
The other parameters are at their reference values.

\begin{figure}
\centering
\includegraphics[angle=-90,width=10cm]{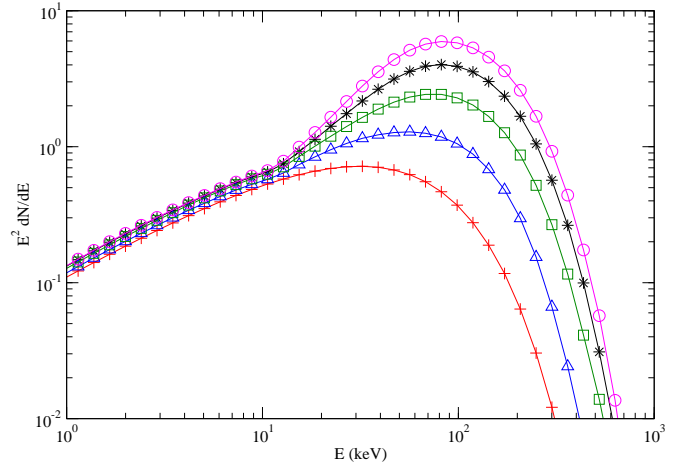}
\caption{
Angle-averaged emergent spectrum in $E^2 dN/dE$ form, with 
arbitrary units, as a function of energy $E$ in keV.  The spectrum 
of the input photons (normalized to unity) is bremsstrahlung (crosses).  
The various curves are the results of different
values of the transverse Thomson optical depth $\tau_{\perp}$:  1 (triagles), 
3 (squares), 5 (stars), and 7 (circles).
The rest of the parameters are at their reference values.
}
\label{Fig2}
\end{figure}

In all cases, a high-energy power-law spectrum with a cutoff is
seen.  The photon-number 
spectral index $\Gamma$ is approximately $1.5$ for 
$\tau_{\perp}=1$, while $\Gamma \approx 0.7$ for $\tau_{\perp}=7$.

\subsubsection{Effects of magnetic-field srength}

As discussed in \S\ 2.2, the larger the cyclotron energy (see eq. 12), the 
smaller the cross section of the low-energy input 
photons of a given energy, and the smaller the
{\it effective} transverse optical depth for the same Thomson transverse 
optical depth $\tau_{\perp}$.  Thus, as the magnetic-field strength
increases, the average number that a photon crisscrosses the shock
decreases and the resulting index $\Gamma$ increases (i.e. steeper
spectrum).  This is clearly seen in Fig. 3,
where the magnetic-field strength is $B=10^{13}$ G and four
values of the Thomson transverse optical depth $\tau_{\perp}$:  
1 (triangles), 3 (squares), 5 (stars), and 7 (circles).
The rest of the parameters are at their reference values.

Comparing Fig. 3 with Fig. 2, we see that the spectra of Fig. 3 
have no break at $\sim 10$ keV and the power laws extend from low 
energies all the way to $E_p \approx 70$ keV.  This is expected, because the
cyclotron energy is $E_c = 116$ keV, which is above $E_p$.
On the other hand, to get the same value of $\Gamma$, one needs a larger
transverse Thomson optical depth in Fig. 3 than in Fig. 2.
For example,
to get $\Gamma \approx 1$, one needs $\tau_{\perp}=7$ in Fig. 3, but only
$\tau_{\perp}=5$ in Fig. 2.  A slight deviation in the value
of the peak energy $E_p$ between Figs. 2 and 3
can be accommodated with a slight change in the
temperature of the post-shock electrons (see \S\ 3.1.3).

\begin{figure}
\centering
\includegraphics[angle=-90,width=10cm]{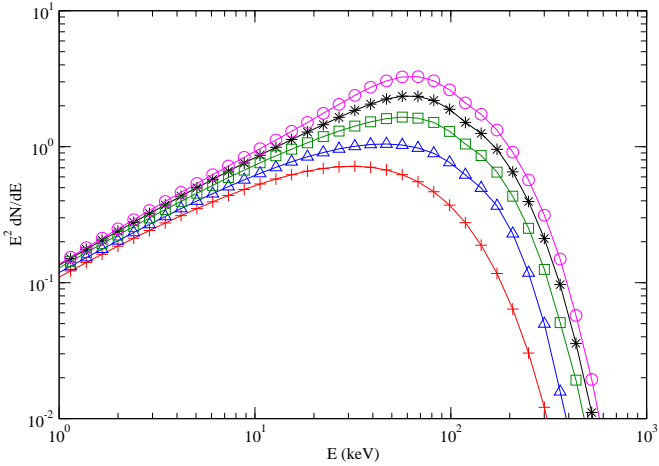}
\caption{
Same as in Fig. 2, but for $B= 10^{13}$ G.
}
\label{Fig3}
\end{figure}

\subsubsection{Effects of post-shock temperature}

As discussed in \S\ 2.1.2, the post-shock temperature is not known.  
We have provided a simple estimate of $kT_e \sim 150$ keV, but 
this is probably an overestimate, because we have assumed a 
one-dimensional Maxwellian distribution.
Thus, we examine two additional values:  100 keV and 
30 keV.

Fig. 4 is similar to Fig. 2, but for $kT_e=100$ keV.  For the same
transverse Thomson optical depth $\tau_{\perp}$, the emergent spectrum is
flatter in Fig. 4 than in Fig. 2.  
For example, a $\tau_{\perp}=3$ is needed in Fig. 4 to produce a
high-energy power-law index $\Gamma \approx 1$, while a 
$\tau_{\perp}=5$ is needed in Fig.2.
Similarly, the peak energy $E_p$ is
larger in Fig. 4 than in Fig. 2.  In particular, for $kT_e = 100$ keV,
the peak energy is $E_p \approx 150$ keV.

Fig. 5 is similar to Figs. 2 and 4, but for $kT_e = 30$ keV.  
The conclusions are the same.  A transverse Thomson optical depth
$\tau_{\perp} = 6$ is required to obtain $\Gamma \approx 1$, while
for $kT_e =50$ keV a $\tau_{\perp} = 5$ is enough.

\begin{figure}
\centering
\includegraphics[angle=-90,width=10cm]{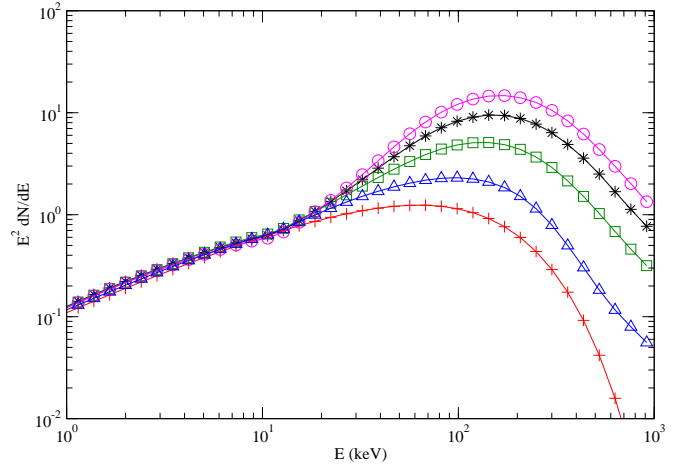}
\caption{
Same as in Fig. 2, but for $kT_e=100$ keV.  
}
\label{Fig4}
\end{figure}

\begin{figure}
\centering
\includegraphics[angle=-90,width=10cm]{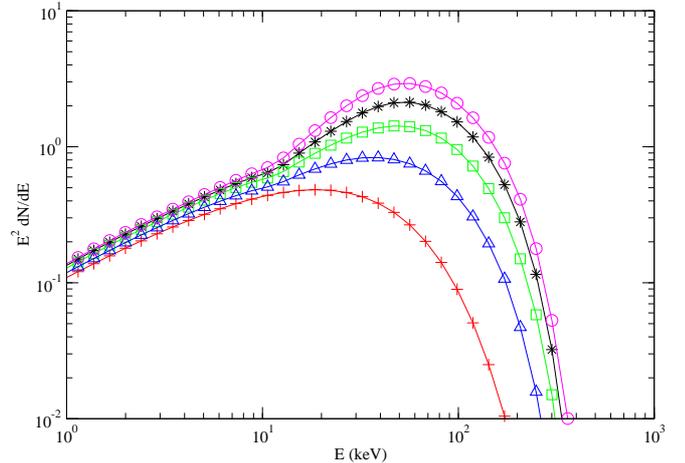}
\caption{
Same as in Fig. 2, but for $kT_e=30$ keV.  
}
\label{Fig5}
\end{figure}

\subsection{Blackbody input}

As discussed extensively in Tr\"umper et al. (2013), in the 
accretion model of AXPs/SGRs the ratio of the observed luminosity $L_s$
of the soft, quasi-thermal photospheric emission to the hard,
fan-beamed, power-law luminosity $L_h$ is of order unity, but it depends
on several parameters, and in particular on the height of the 
radiative shock.  In other words, the larger the height, the larger the
radius $r_0$ of the polar cap.  This is true not only for geometrical 
reasons but also due to gravitational bending.  

Since $r_0 \sim 8$ km for 4U 0142+61 (White et al. 1996; 
Israel et al. 1999; Juett et al. 2002), while the radius of the
radiative shock is typically $a_0 \sim 100$ m (see eqs. 13 and 14),
we estimate that only a small fraction 
($\sim 10^{-4}$, see \S\ 3.3) of the photospheric
photons enter the shock.  The majority of them come from around the 
footprint of the accretion column.  We make the approximation that they are 
described by a blackbody distribution of temperature $T_{BB}$ and that
they enter the shock at an angle $\theta$ with respect to the axis of the 
accretion column.  For simplicity we take $\cos \theta =0.7$.
Our results are insensitive to this last approximation.

For a representative calculation with blackbody input,
we use the following values of the parameters:
$\tau_{\perp}= 5$, $kT_e= 50$
keV, $B=10^{12}$ G, and $kT_{BB}=0.5$ keV.

In Fig. 6, we show the input blackbody spectrum (crosses),
with the integral of $dN/dE$ normalized to unity,
and the resulting angle-averaged output
spectrum of the Monte carlo (stars) for the
above values of the parameters.  A hint of a 
broad peak at low energies ($E \simless 10$ keV) 
is due to the input photons that 
escape unscattered or suffer
a small number of scatterings, and can be thought of as a ``shift'' of the 
blackbody spectrum to higher energies,
though it should be kept in mind that what is plotted is
$E^2 dN/dE$ and not $dN/dE$.  

Above about 10 keV, the spectrum is a power law, with photon number
spectral index $\Gamma \approx 1.3$ and a peak in the $E^2 dN/dE$
distribution at $E_p \approx 80$ keV.
This is again a natural consequence of first order Fermi energization.
Despite the fact that only a small number of photons suffer many
crisscrosses of the shock, most of the energy in the emergent spectrum
comes out at $E_p$.  

\begin{figure}
\centering
\includegraphics[angle=-90,width=10cm]{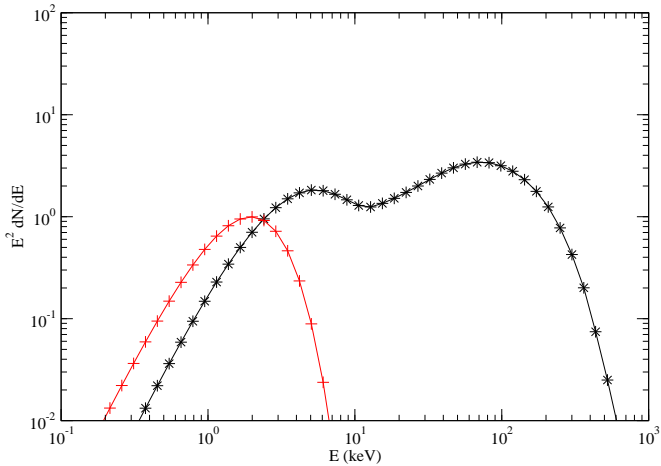}
\caption{
Angle-averaged emergent spectrum (stars) in $E^2 dN/dE$ form, with 
arbitrary units, as a function of energy $E$ in keV.  The spectrum 
of the input photons (normalized to unity) is blackbody (crosses).  
The parameters are:
$\tau_{\perp}= 5$, $kT_e= 50$ keV, $B=10^{12}$ G, and $kT_{BB}=0.5$ keV.
}
\label{Fig6}
\end{figure}

Comparing Figs. 1 and 6, we see that for the same transverse
optical depth $\tau_{\perp} = 5$ the photon number spectral index 
$\Gamma$ is approximately equal to 1 in Fig. 1 and approximately 1.3
in Fig. 6.  This means that bremsstrahlung is more efficient in
producing the high-energy power law than the blackbody.  This is, of course,
natural because the bremsstrahlung contains photons with energy
comparable to $kT_e$, while the typical blackbody-photon energy is
$2 kT_{BB} << kT_e$. Thus, fewer scatterings are needed 
for some bremsstrahlung photons to reach the 
peak energy $E_p \approx 80$ keV than for all the blackbody ones.

\subsection{Blackbody versus bremsstrahlung input}

Despite the fact that the quasi-thermal, photospheric
photons can in principle produce high-energy power-law spectra
similar to the ones oberved in AXPs/SGRs by Comptonization in
the radiative shock (e.g., Fig. 6), we will show
below that their role is rather insignificant compared to that of the
bremsstrahlung ones.  The reasons are the following:

1) Most of the photospheric photons do not enter the shock.  For simplicity
we assume that the polar cap is flat.  This does not introduce any 
significant error, because gravitational bending applies to both the
hard X-rays from the shock to the polar cap and to the soft X-rays
from the polar cap to the shock.  Let the area of the shock seen by
a photospheric photon be 
$A \approx \bar l_e 2 a_0 = 2 a_0^2/\tau_{\perp} = 0.4 a_0^2$ for 
$\tau_{\perp}=5$, 
where $\bar l_e = a_0/\tau_{\perp}$ is the shock thickness (see eq. 3).
The fractional solid angle subtended by the area $A$ at radius $r$ 
of the polar cap is $d\Omega / 4 \pi = A/4\pi r^2$ and for isotropic
emission from the polar cap we find that the fraction of polar cap
photons that reach the shock is 
$$
f_{BB} \approx { {A} \over {2 \pi r_0^2} } \ln { {r_0} \over {a_0} } 
\approx 10^{-4},
\eqno(21)
$$
for $r_0 = 8$ km and $a_0 = 100$ m.

2) The number of photospheric photons that enter the radiative shock 
is smaller than the corresponding bremsstrahlung ones.  This is because,
for accretional luminosity $L$, the photospheric luminosity that enters
the shock is about $10^{-4} (L/2)$,
while the corresponding bremsstrahlung luminosity, i.e., the 
remaining luminosity in the post-shock region is 
(see \S\ 2.1.2) $\sim (m_e/m_p)L$.  For a typical photon energy of 1 keV for 
the photospheric emission and less than 1 keV for
bremsstrahlung, we see that 
bremsstrahlung dominates the soft photon input to the shock.

3) As discussed in \S\ 3.2, for equal inputs to the radiative shock, 
the bremsstrahlung is more effectively upscattered to the high-energy
power law than the photospheric input.

\section{Discussion}

It is known observationally (e.g., Enoto et al. 2010) that in
AXPs/SGRs the hard X-ray luminosity $L_h$ is comparable to the soft
X-ray one $L_s$.  In the accretion picture presented by Tr\"umper
et al. (2010, 2013), this is a natural consequence of the fact that only
about half of the produced hard X-ray luminosity 
in the radiative shock is observed directly 
($L_h$), while the other half is intercepted by the polar cap and the 
major fraction of it is re-emitted as soft X-rays ($L_s$).

With our reported calculations in \S\ 3.1, we have shown that the radiative
shock can produce a fan-beam, hard X-ray, power-law spectrum similar to that
seen in AXPs/SGRs.  Furthermore, in \S\ 3.3 we have shown that nearly all
the photospheric luminosity $L_s$ is observed unaltered.  Thus,
the total X-ray spectrum, soft and hard, of an AXP/SGR with $L_h = L_s$
can be easily produced by adding to the hard spectrum shown in Figs. 1 - 5
a blackbody spectrum with a peak in $E^2 dN/dE$ having the same height
as the peak of the hard X-ray spectrum.  

As an example, we consider the hard X-ray spectrum of Fig. 1 (stars)
and add to it a blackbody spectrum of temperature $T_{BB}=0.5$ keV that has 
the same height in $E^2 dN/dE$ as the hard X-ray spectrum.  The result is 
shown in Fig. 7.

In view of the above, it is natural to expect that low-luminosity X-ray pulsars 
should exhibit soft and hard X-ray spectra similar to those seen in AXPs and
SGRs.  Indeed, as discussed in Tr\"umper et al. (2013), 4U 0352+309 (X-Per) 
and 4U 2206+54 show high-energy power-law tails extending to $\sim 100$ keV,
with photon number spectral index $\Gamma \sim 1$.

\begin{figure}
\centering
\includegraphics[angle=-90,width=10cm]{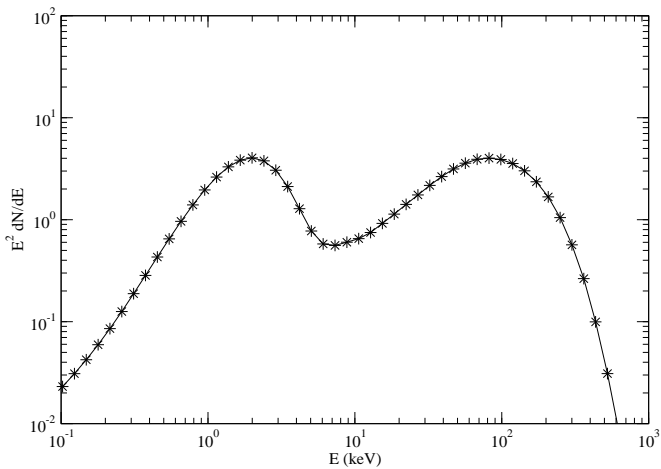}
\caption{
Characteristic spectrum expected from low-luminosity accretion onto 
a magnetic neutron star.  The high-energy spectrum is produced in the
radiative shock near the bottom of the accretion column.  The parameters
are:  $\tau_{\perp}=5$, $kT_e=50$ keV, and $B=10^{12}$ G.
We have added to it a blackbody spectrum of temperature
$kT_{BB} = 0.5$ keV such that its peak has the same height as
the peak at 80 keV.
}
\label{Fig7}
\end{figure}

\section{Conclusions and predictions}

We have demostrated with a simple quantitative model 
that accretion from a fallback disk onto 
a neutron star with normal ($10^{12} - 10^{13}$ G) dipole magnetic field can
produce hard X-ray, power-law spectra similar to the ones observed from 
steady-state or transient AXPs and SGRs.
Furthermore, we have shown that the soft part of the X-ray spectrum,
that comes from the polar cap, does not enter in a significant way
in the formation of the hard part of the spectrum and can simply be 
added to it.

The power-law index $\Gamma$ of the hard X-ray spectrum depends mainly 
on the accretion rate, or equivalently on the transverse optical depth
of the accretion column at the radiative shock, and less on the strength
of the magnetic field and the post-shock temperature.  A 
transverse Thomson optical depth of
$\sim 5$ is enough to produce a high-energy power-law index 
$\Gamma \approx 1$.

The cutoff at high energies ($E \sim 100$ keV), is due to the 
limited kinetic energy of the accreting electrons, and it depends 
on the post-shock temperature.  
If the accretion rate is sufficient for the formation of a radiative shock, 
the cutoff energy 
cannot be larger than $\sim 300$ keV.  
Thus, if our picture is correct, we predict that {\it no AXP or SGR will be 
found to have a cutoff energy of 400 keV or larger
in their steady-state or transient spectra}.

\begin{acknowledgements}

We thank an anonymous referee for useful comments, which have 
improved our paper. 
One of us (NDK) acknowledges partial support by the ``RoboPol" project, 
which is 
implemented under the ``ARISTEIA" Action of the ``OPERATIONAL PROGRAM 
EDUCATION AND LIFELONG LEARNING" and is co-funded by the European Social
Fund (ESF) and National Resources.
NDK also acknowledges a grant from the European Astronomical Society 
in 2013.
\"{U}. E. acknowledges research support from
T\"{U}B{\.I}TAK (The Scientific and Technical Research Council of Turkey) 
through grant 113F166.
\end{acknowledgements}

\end{document}